\documentclass[aps,prb,twocolumn]{revtex4}
\usepackage{bm}
\usepackage{graphicx}

\begin{document}
\title{Comment on ``Influence Functional for Decoherence of Interacting
  Electrons in Disordered Conductors'' (cond-mat/0510563v1) and on related
  papers (cond-mat/0510556v1 and cond-mat/0510557v1)}
\author{Dmitri S. Golubev$^{1,3}$ and Andrei D. Zaikin$^{2,3}$}
\affiliation{$^1$Institut f\"ur Theoretische Festk\"orperphysik,
Universit\"at Karlsruhe, 76128 Karlsruhe, Germany \\
$^2$Forschungszentrum Karlsruhe, Institut f\"ur Nanotechnologie,
76021 Karlsruhe, Germany\\
$^3$I.E.Tamm Department of Theoretical Physics, P.N.Lebedev
Physics Institute, 119991 Moscow, Russia}

\begin{abstract}
Recently von Delft \cite{vD1} (JvD) has successfully re-derived
our influence functional for interacting electrons and claimed
that within our approach he was able to obtain the electron
decoherence rate that vanishes at $T=0$. In this Comment we
demonstrate that this JvD's claim is in error, as it is based on
ambiguous and uncontrolled manipulations violating basic
principles of quantum theory, such as energy-time uncertainty
relation, causality, fluctuation-dissipation theorem, detailed
balance and the like. We also briefly address insufficient
approximations employed by Marquardt {\it et al.} \cite{vD2} and
by von Delft {\it et al.} \cite{vD3} and demonstrate that the
results of all three papers \cite{vD1,vD2,vD3} in the limit $T \to
0$ are inconsistent with simple rules of algebra.

\end{abstract}

\maketitle

In recent papers \cite{vD1,vD2,vD3} von Delft (JvD) {\it et al.}
addressed the problem of low temperature decoherence of electrons
in disordered conductors. JvD \cite{vD1} has re-derived our
influence functional approach for interacting electrons in
disordered conductors \cite{GZ2,GZ3,GZS}, observed that it
``properly incorporates the Pauli principle'' and acknowledged
that we ``got it completely right'' which is ``a significant and
important achievement''. At the same time, JvD \cite{vD1} claimed
that -- in comparison to our work -- he has achieved ``a more
accurate treatment of recoil effects''. According to JvD
\cite{vD1}, ``with this change'' he could reproduce ``in a
remarkably simple way, the standard, generally accepted results
for the decoherence rate''.

Marquardt  {\it et al.} (MDSA) \cite{vD2,vD3} re-iterated the same
conclusions. On one hand, they also admitted that we
\cite{GZ2,GZ3,GZS} ``proposed a new, exact Feynman-Vernon
influence functional for electrons'' which ``takes proper account
of the Pauli principle''. On the other hand, MDSA confirmed JvD's
claim \cite{vD1} ``that if the Pauli blocking terms are treated
somewhat more carefully to include recoil effects'' our influence
functional ``approach actually does reproduce'' the decoherence
rate ``which does not saturate at low temperatures''.

To summarize, JvD {\it et al.} \cite{vD1,vD2,vD3} make two key
statements: (i) our influence functional \cite{GZ2,GZ3,GZS} is
correct and {\bf exact} \cite{FN1} and (ii) our calculation of the
corresponding path integral is not correct as it ``unintentionally
neglects recoil effects''.

The statement (i) is very important. After numerous unsuccessful
attempts (reviewed, e.g., in Refs. \onlinecite{GZS,GZ4,vD1}) to
search for a mistake in our derivation of the influence functional
\cite{GZ2,GZ3,GZS} it is finally acknowledged by JvD and MDSA that
this mistake just does not exist. Note that previously our
influence functional approach has been independently re-derived by
Eriksen {\it et al.} \cite{Er} who also confirmed its validity.
The observation (i) allows to bypass practically all steps in our
derivation restricting the whole discussion to just one -- purely
mathematical -- issue, i.e. how to correctly evaluate our path
integral \cite{GZ2}.

In this Comment we demonstrate that JvD's analysis of our
influence functional and his statement (ii) about recoil effects
``neglected'' in our calculation are in error. Instead of directly
evaluating our influence functional JvD effectively replaces it
by a very different ``influence functional for decoherence''
obtained by performing uncontrolled manipulations with diagrams.
As a result JvD \cite{vD1} arrives at the expression for this 
functional which violates fundamental principles of quantum theory. In
particular, by dropping certain classes of diagrams JvD violates
analytic properties of the influence functional \cite{FH} which in turn 
implies {\it
violation of causality}. By ambiguously suppressing fluctuations
of energy of interacting electrons JvD {\it violates energy-time
uncertainty relation}. By selectively neglecting photon
frequencies in infinite series of diagrams JvD effectively breaks
down thermal equilibrium in the photon subsystem and, hence, {\it
violates fluctuation-dissipation theorem (FDT) and detailed
balance.}

The structure of our Comment is as follows.  In Sec. I we inspect
JvD's  ``influence functional for decoherence'' and demonstrate
its inconsistency  with basic principles of quantum theory. In
Sec. II we explicitly analyze the main drawbacks of JvD analysis
which have eventually led to the above problems. In Sec. III we
briefly discuss insufficient approximations used by MDSA
\cite{vD2,vD3} and conclude the discussion by demonstrating that
both JvD's and MDSA's results for zero temperature electron
decoherence rate can be discarded without any calculation just on
the basis of simple rules of algebra. This part of our discussion
is easily accessible also to non-experts.

\section{Some consequences of JvD's analysis}

The central result of our derivation \cite{GZ2} is a general
expression for the conductivity of a disordered conductor in terms
of the Feynman-Vernon influence functional for interacting
electrons. The corresponding expressions are presented in details
in our earlier publications and briefly summarized  in Appendix A
for convenience.

The key role in our analysis is played by the path integral
\begin{eqnarray}
J=\int{\cal D}^2{\bm p} \int{\cal D}^2{\bm r}{\rm
e}^{\frac{i}{\hbar}(S_0[ {\bm p}_{F},{\bm r}_{F}] -S_0[{\bm
p}_{B},{\bm r}_{B}])}F, \label{J0}
\end{eqnarray}
which defines the kernel $J$ of the evolution operator on the
Keldysh contour. Here  $S_0$ is the action for non-interacting
electrons (\ref{S0}) and
\begin{eqnarray} F={\rm e}^{-(1/\hbar)(iS_{
R}[{\bm p}_{F,B},{\bm r}_{F,B}] + S_{I}[{\bm r}_{F,B}])}
\label{F0}
\end{eqnarray}
is the electron influence functional. It is important that the
term $S_R$ (\ref{SR}) contains the electron density matrix $\rho$
which makes it necessary to keep integrals over both coordinate
and momentum variables because the corresponding Hamiltonians turn
out to be nonlocal. It is also important to observe that both
dissipative term $S_R$ and noise term $S_I$ are {\it purely real} functionals.

In order to describe weak localization in the limit $k_Fl \gg 1$
it is sufficient to employ the standard picture of time-reversed
paths and evaluate the path integral (\ref{J}) within the saddle
point approximation. It turns out that the term $S_R$ vanishes for
any pair of time-reversed paths \cite{FN3} while $S_I$ grows with
time at all temperatures including $T=0$ and determines the
Cooperon decay at sufficiently long times. One can also evaluate
the contribution of fluctuations around time-reversed paths which
turns out to be significant only at short times, i.e. in the
perturbative (in the interaction) regime. The corresponding
analysis was carried out in Ref. \onlinecite{GZ4}. Employing an
exact transformation of the evolution operators we reduced the
result to the form allowing to semiclassically evaluate only those
path integrals which do not contain the electron density matrix
$\rho$ while the latter was kept in its exact form in the
corresponding matrix elements which cannot grow with time. This
analysis was performed within the accuracy of the {\it definition}
of the weak localization correction to conductivity and fully
confirmed our earlier results and conclusions.

Unfortunately this latter development was completely ignored by
JvD who attempted to treat the path integral yet ``somewhat more
accurately'' than we do. JvD's key idea is to effectively
integrate out the momentum variables in Eq. (\ref{J}) which would
yield the ``position-only'' representation of the influence
functional. Instead of (\ref{J}) JvD obtained:
\begin{equation}
\int{\cal D}^2{\bm r} {\rm e}^{\frac{i}{\hbar}(\tilde S_0[{\bm
r}_{F}] -\tilde S_0[{\bm r}_{B}])}\tilde F_\epsilon , \label{vDJ}
\end{equation}
where $\tilde S_0[{\bm r}]$ is the action for non-interacting
electrons and
\begin{equation} \tilde F_\epsilon =
{\rm e}^{-(1/\hbar)(i\tilde S^{\epsilon}_R[{\bm r}_{F,B}] + 
S_I[{\bm r}_{F,B}])}, \label{FvD}
\end{equation}
where
$\tilde S_R^\epsilon$ is defined in Eq. (\ref{tildeSR}).

Below we will demonstrate that the functional (\ref{FvD}) is entirely
different from the correct one (\ref{F0}) and it cannot be obtained from the
latter by performing momentum integrations or by any other correct means.
 
A quick glance already at the first JvD's formula (Eq. (1a,b) in Ref.
\onlinecite{vD1}) demonstrates that this and subsequent equations
violate the energy-time uncertainty relation. Indeed, on p. 5 of
Ref. \onlinecite{vD1} JvD writes that the double path integral
entering his Eq. (1b) ``gives the amplitude for an electron with
energy $\hbar \epsilon$ to propagate from $r_{2'}$ at time $-\tau
/2$ to $r_1$ at $\tau /2$ times the amplitude for it to propagate
from $r_{1'}$ at time $\tau /2$ to $r_2$ at $-\tau /2$''. Thus, in
JvD's Eq. (1) and his subsequent analysis both electron {\it
energy and time} are simultaneously fixed. This is a clear
{\bf violation of the quantum mechanical energy-time uncertainty
relation}. No such amplitude can be defined in quantum mechanics.

Let us further examine the ``influence functional'' defined in
Eqs. (2-4) of Ref. \onlinecite{vD1}. It is easy to observe that,
while our action $S_R$ (\ref{SR}) is purely real, JvD's action
$\tilde S_R^\epsilon$ already contains an imaginary part. Since
this part effectively adds up to the noise term $S_I$ it would 
imply that the equilibrium Nyquist noise spectrum ``felt'' by electrons
in a disordered conductor would be proportional to the combination
$\coth\frac{\omega}{2T}+\frac12(\tanh\frac{\epsilon -\omega}{2T}-\tanh\frac{\epsilon +\omega}{2T})$
and not to the commonly accepted $\coth\frac{\omega}{2T}$ which also
follows from our expression for $S_I$ (\ref{SI}), (\ref{RI}). Already
this observation demonstrates that JvD's ``influence functional'' is
fundamentally different from ours. While the latter takes full account
of both classical and quantum noise, JvD's expressions
(\ref{tildeSR}-\ref{imR}) would imply ``cutting out'' the quantum noise 
for small values of $\epsilon$. As a result, according to JvD in the limit
$\epsilon =0$ and $T \to 0$ electrons would ``feel'' no noise at all.
In this limit JvD's
$i\tilde S_R$ {\it exactly} cancels $S_I$, i.e. the ``influence
functional'' $\tilde F_{\epsilon =0}$ (\ref{FvD}) is identically equal to unity
for {\it all} electron paths. In other words, JvD's analysis
predicts that electron-electron
interactions would have {\it no influence} at all on the dc
conductivity of a disordered conductor at $T=0$. This is obviously
not the case for our influence functional \cite{GZ2,GZ3,GZS}.

Our next observation is that JvD's ``influence functional'' {\bf
violates causality}. This fact becomes obvious by inspection of
analytic properties of the Fourier-transformed JvD's function
$\tilde R_\epsilon$ (\ref{reR},\ref{imR}) which has poles both in the
upper and lower half-planes of the $\omega$-variable, hence,
implying that the electron motion should be affected by photons
coming both from the past and from the future, an obvious
nonsense.

In contrast, any correct expression for the kernel $R (\omega
,{\bm q})$ should have poles only in the lower half-plane of
$\omega$, which is the case for our expression
(\ref{RI}-\ref{epsilon}). Applying the 
least action conditions to our action
\begin{eqnarray}
\frac{\delta\big(S_0[{\bm p}_F,{\bm r}_F]-S_0[{\bm p}_B,{\bm r}_B]-S_R[{\bm p}_{F,B},{\bm r}_{F,B}]\big)}
{\left.\delta ({\bm r}_F-{\bm r}_B)\right|_{{\bm r}_F={\bm r}_B={\bm r},{\bm p}_F={\bm p}_B={\bm p}}}=0,
\label{dSdr}\\
\frac{\delta\big(S_0[{\bm p}_F,{\bm r}_F]-S_0[{\bm p}_B,{\bm r}_B]-S_R[{\bm p}_{F,B},{\bm r}_{F,B}]\big)}
{\left.\delta ({\bm p}_F-{\bm p}_B)\right|_{{\bm r}_F={\bm r}_B={\bm r},{\bm p}_F={\bm p}_B={\bm p}}}=0,
\nonumber
\end{eqnarray}
one recovers the standard {\it classical} 
equation of motion for a high energy
electron in a dissipative environment \cite{LL}
\begin{equation}
m\ddot{\bm r}+\nabla U(\bm{r})+e^2\int_{-\infty}^t dt' \nabla
R(t-t',\bm{r}(t)-\bm{r}(t'))=0, \label{cl1}
\end{equation}
where the time integral in a dissipative term extends 
from $-\infty$ to $t$ in full
agreement with the causality principle. In contrast, the same
equation derived from JvD's action (\ref{Pe},\ref{tildeSR}) would take
the form
\begin{equation} m\ddot{\bm r}+\nabla
U(\bm{r})+e^2\int\limits_{-\infty}^{+\infty} dt' \nabla
\tilde R_\epsilon^{Re}(t-t',\bm{r}(t)-\bm{r}(t'))=0,
\label{cl2}
\end{equation}
where the time integral already runs over all times
between $-\infty$ and $+\infty$ in a direct conflict with the
requirement of causality. In addition, instead of the correct
kernel $R(t,\bm{r})$ (\ref{RI}) JvD's Eq. (\ref{cl2}) 
contains $\tilde R_\epsilon^{Re}
(t,\bm{r})$ (\ref{reR}) in a clear contradiction to the 
well established results \cite{LL}. Thus, JvD's ``influence
functional'' fails already on a {\it classical} level being unable
to correctly describe particle's dynamics in a dissipative
environment.

 In addition to the above problems
 JvD's ``influence functional'' also {\bf
violates fluctuation-dissipation theorem and detailed balance}.
The easiest way to observe this violation is to consider the
values of $\epsilon$ large compared to temperature $T$. In that
case the Pauli principle should not be important and the term
$\tilde S_R^\epsilon$ should approach the action $S_R$ evaluated
without the Pauli principle, e.g. for a muon moving in a metal.
This situation was considered, e.g., in Ref. \onlinecite{GSZ} and
the corresponding expression for $S_R$ obtained there coincides
with (\ref{SR}) where one should now set $\rho$ equal to zero. In
this case FDT is represented by the following relation
\begin{equation}
I(\omega , {\bm q})=-\coth \frac{\omega}{2T}{\rm Im}R(\omega , {\bm q}).
\label{FDT}
\end{equation}
While FDT (\ref{FDT}) is manifestly satisfied for our influence
functional, JvD's kernel $\tilde R_\epsilon (\omega ,{\bm q})$
(\ref{reR},\ref{imR}), being substituted into Eq. (\ref{FDT})  instead of
$R$, obviously violates this relation. As a direct consequence,
JvD's ``influence functional'' also violates detailed balance.

Concluding this section, we have demonstrated that there exists no
environment in Nature which could be described by the ``influence
functional'' defined in Eqs. (2-4) of Ref. \onlinecite{vD1}.
Hence, this ``influence functional'' is unsuitable for
calculations of any physical quantity including, of course, the
electron decoherence rate.

\section{Main problems of JvD's analysis}

We now turn to the main problems of JvD's calculation which have
eventually led to alarming consequences outlined above.

As it was already emphasized, the path integral over momentum
variables (Eqs. (\ref{J}),(\ref{S0}),(\ref{SR})) is strongly
non-Gaussian, and, hence, cannot be exactly evaluated. JvD
\cite{vD1} does not even attempt to directly perform the momentum
integrals. Instead, he carries out a set of manipulations and
formulates ``a rule of thumb'', Eq. (B.91). Unfortunately this
central part for the whole analysis (pp. 54-55 of the paper) is by
far less detailed than the rest of the paper and, in fact,
contains almost no equal signs (substituted in eqs. (B.91-B.93) by
the sign ``$\to$'').

What was actually done by JvD with our influence functional?  JvD
uses it only at the very first step expanding $F$ in Eqs. 
(\ref{J0},\ref{F0}) in
powers of $iS_R+S_I$. In this way JvD reproduces all
Keldysh diagrams for the problem in question. This step is correct
and -- in full agreement with our statements
\cite{GZ2,GZ3,GZS,GZ4} -- demonstrates that no diagrams are
missing within our influence functional approach. From this point
on JvD performs his own analysis of the diagrammatic series. This
analysis has no direct connection to any path integral
calculation, since all Keldysh diagrams can, of course, be also
recovered without path integrals. JvD's key steps are as follows:

\begin{enumerate}

\item Rewriting the Keldysh Green function for electrons as
$G^K(E-\omega)=[G^R(E-\omega)-G^A(E-\omega)]\tanh\frac{E-\omega}{2T}$
JvD neglects $G^A$ on the forward branch of the Keldysh contour, i.e. he makes the replacement
$$G^K(E-\omega)\to G^R(E-\omega)\tanh\frac{E-\omega}{2T},$$
see his Eq. (B.93). Similarly, on the backward branch of the
Keldysh contour JvD replaces
$$G^K(E-\omega)\to -G^A(E-\omega)\tanh\frac{E-\omega}{2T}.$$
This approximation is equivalent to simply dropping certain
classes of diagrams. Already at this point JvD violates causality
for electrons.

\item JvD splits the Pauli factor $\tanh\frac{E-\omega}{2T}$ from
$G^K$ (or, better to say, from the remaining part of $G^K$
containing $G^R$ or $G^A$ only) and transfers it to the photon
propagators ${\cal L}^{R}$ on the forward or ${\cal L}^A$  on the
backward branches of the Keldysh contour, i.e. he makes the
following replacement
\begin{eqnarray}
&& G^K(E-\omega){\cal L}^{R,A}(\omega)\to
\nonumber\\
&& \to
\pm G^{R,A}(E-\omega)\;\tanh\frac{\epsilon-\omega}{2T}{\cal
L}^{R,A}(\omega).
\nonumber
\end{eqnarray}
 Note, that the energy $\epsilon$ under $\tanh$
is now different from $E$. According to JvD, $\epsilon$ is set to
be constant which is not sensitive to the pre-history. At this
stage JvD violates the energy-time uncertainty relation as well as
causality for photons. This step in combination with the previous
one is equivalent to JvD's ``rule of thumb'', Eq. (B.91), applied
to the first order diagrams.

\item JvD spreads his ``rule of thumb'' to all orders of the
perturbation theory. For that purpose in addition to steps (1) and
(2) JvD  substitutes
$$\tanh\frac{E-\omega_1-\omega_2-...-\omega_n}{2T}\to
\tanh\frac{\epsilon-\omega_1}{2T}
$$
in all the diagrams whenever more than one photon frequency under
$\tanh$ is encountered. As a result, all the complicated Pauli
factors are reduced to a simple and unique form
$\tanh\frac{\epsilon-\omega }{2T}$ for {\it all diagrams in all
orders}. Hence, all tanh-factors become decoupled from the
electron lines and can now be absorbed in the photon Green
functions. After this step electrons are not anymore sensitive to
the Pauli principle. In addition, this manipulation leads to
violation of FDT, since it effectively breaks down thermal
equilibrium in the photon subsystem.

\end{enumerate}

\begin{figure}
\includegraphics[width=9cm]{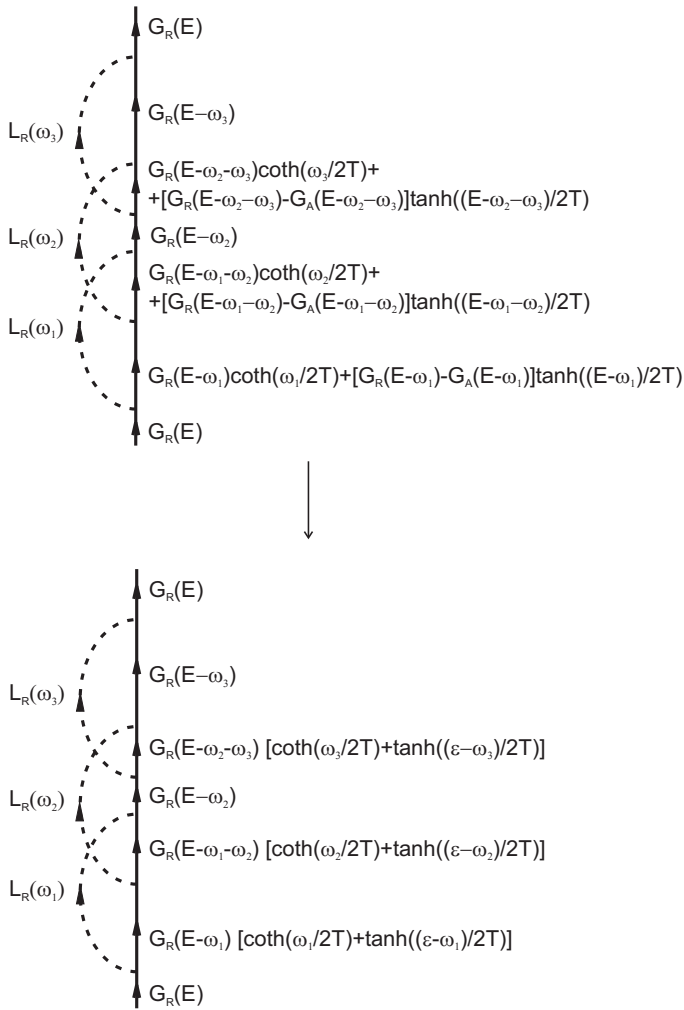}
\caption{JvD's ``rule of thumb'' illustrated for one of the third
order diagrams. All three steps (1) to (3) are evident. Notice
that the Golden rule combination
$\coth\frac{\omega}{2T}+\tanh\frac{\epsilon-\omega}{2T}$ appears
only after JvD's ``rule of thumb'' is applied, while the initial
exact expression for this diagram is by far more complicated and
it does not contain this simple combination of coth and tanh.}
\end{figure}

Summarizing, the above set of steps is equivalent to (a) dropping
certain classes of diagrams and (b) replacing infinitely many
remaining diagrams by completely different ones. JvD's ``rule of
thumb'' is also illustrated in Fig. 1.

Unfortunately no serious justification for these manipulations was
offered by JvD. For instance, according to JvD \cite{vD1}, some
contributions to the Cooperon self-energy corresponding to the
so-called Hikami boxes \cite{FN4} should vanish after impurity
averaging in the first order in the limit of zero frequency and
wave vectors. For some unclear reasons JvD believes that this
observation should be sufficient in order to perform his step (1),
i.e. to disregard terms with $G^A$ and $G^R$, already {\it before}
impurity averaging and {\it in all orders} of the perturbation
theory.

As for the step (2), according to JvD it is possible to neglect
all energy fluctuations ``if one so chooses''. This approximation
``is expected to work well if the relevant physics is dominated by
low frequencies''. This argument is logically inconsistent, since
in the very beginning it already {\it assumes} what one needs to
{\it prove} in the end. In practice, as it is argued by JvD in
Sec. B6.2, one should disregard ``accumulation of energy changes''
in all vertex diagrams in all orders and also in the self-energy
diagrams in the second and all higher orders in the interaction.
Our analysis, in contrast, demonstrates that this ``energy
accumulation'' is important and cannot be disregarded.

The combination of JvD's steps (1) and (2) makes it impossible to
fully reconstruct all contributions to the Cooperon already in the
first order in the interaction. For instance, the first order
non-Golden-rule terms defined, e.g., in Eqs. (70) of Ref.
\onlinecite{GZ3} are completely missing in JvD's Eqs. (1-4) and
his subsequent analysis. The time dependence encoded in these
non-Golden-rule terms is indeed slower than linear. Nevertheless,
these first order terms are important at $T \to 0$ since in 1d and
2d cases they actually {\it diverge} in the long time limit and,
hence, also contribute to decoherence \cite{FN2}, see Sec. IVB of
Ref. \onlinecite{GZ3}.

JvD's step (3) introduces yet one more uncontrolled approximation
by neglecting all but one photon frequencies under each tanh. To
support this step JvD argues that all these frequencies are
smaller than $T$.  It is worth stressing that in the exact
diagrammatic expansion the photon frequencies are {\it not}
restricted by temperature. The contribution of high frequencies
is important and may even lead to ultra-violet divergencies. Only after the
steps (1) and (2) such divergencies could disappear. However, even
if one adopts these steps, it would still be inconsistent to
disregard all but one photon frequency. Since all these
frequencies can be of the same order, it is simply a matter of
consistency of any approximation to either keep or disregard them
all. In either case it would then be impossible to recover the
desired combination
$\coth\frac{\omega}{2T}+\tanh\frac{\epsilon-\omega}{2T}$. Perhaps,
this observation might suggest a clue why only one photon
frequency was kept by JvD under each tanh.

The net result of all these manipulations is nothing but simple
exponentiation of the first order self-energy diagrams \cite{FN6}
evaluated within the approximation which effectively ignores all
contributions not containing the combination
$\coth\frac{\omega}{2T}+\tanh\frac{\epsilon-\omega}{2T}$. For the
latter reason already the first order terms of the perturbation
series cannot be fully recovered from JvD's Eqs. (1-4). However,
even if all the first order terms would be correctly reproduced by
JvD, it would still be completely useless to simply exponentiate
the first order result. The problem in question is essentially
non-perturbative and, hence, contributions of all orders should be
fully included \cite{GZ2,GZ3,GZS,GZ4}.

After all these manipulations with diagrams JvD goes back to path
integrals and constructs the functional (\ref{FvD}) which, being
expanded in the interaction, should generate JvD's diagrams
obtained after his steps (1) to (3). Obviously, this functional
has nothing to do with our influence functional (\ref{F0}) no
matter whether one carries out momentum integrations or not. In
fact, JvD himself acknowledges that ``an accurate treatment of
effects occurring in second or higher order is beyond the accuracy
of'' his ``influence functional approach''. In contrast to
(\ref{FvD}), our influence functional (\ref{F0}) -- as was also
confirmed by JvD -- includes all RPA diagrams in all orders and,
hence, is suitable for non-perturbative calculations. Thus, the
crucial  difference between (\ref{F0}) and (\ref{FvD}) becomes
evident already by careful reading of last paragraphs of Sec. B6.2
of Ref. \onlinecite{vD1}.

Unfortunately in other parts of that paper the wording is
sometimes not so clear. For instance, in Sec. 3 on p. 9 JvD first
acknowledges that ``we got it completely right in the
position-time representation'' and then continues ``unfortunately,
however, it did not occur to them to use the frequency
representation (4e)''. Having read these statements, the reader
could easily conclude that JvD only passes to a different
representation of {\it the same} influence functional. In reality,
however, at this point JvD {\it replaces} our correct influence
functional (\ref{F0}) by a very different object (\ref{FvD}) which
properties have already been discussed in Sec. I of this Comment.

In Sec. 4 of Ref. \onlinecite{vD1} JvD incorrectly ascribes to us
certain manipulations with {\it his} functional (\ref{FvD}) which
we have never performed and could never perform simply because we
do not have such a functional at all in any of our papers. The
procedure discussed in Sec. 4 of Ref. \onlinecite{vD1} has,
therefore, nothing to do with our saddle point analysis
\cite{GZ2,GZ3,GZS} and JvD's claim that this analysis ``neglects
recoil effects'' is highly misleading. Obviously, all recoil
effects  are fully included in our path integral
(\ref{J0})-(\ref{F0}). Evaluation of this non-Gaussian path
integral within the saddle point approximation is a legitimate and
standard mathematical procedure. It is certainly correct within
its applicability range and by no means it implies neglecting
recoil effects, as it was incorrectly alleged by JvD.

\section{Further remarks and algebraic considerations}
For completeness, let us briefly address MDSA papers
\cite{vD2,vD3}. The analysis in Sec. VII of Ref. \onlinecite{vD2}
is essentially identical to that of Ref. \onlinecite{vD1} and,
hence, all our critique of the latter paper equally applies to
MDSA's work \cite{vD2}. In particular, the action defined in Eqs.
(93)-(94) of Ref. \onlinecite{vD2} as well as MDSA's
``plausibility arguments'' also violate fundamental principles of
quantum theory as already discussed in Sec. I of this Comment.
MDSA's  statement that our effective action $iS_R+S_I$ ``is
essentially the same as'' that defined in their Eqs. (93)-(94) is
explicitly incorrect, as it was already argued above. These two
actions are so different, that they even yield different {\it
classical} equations of motion, respectively Eqs. (\ref{cl1}) and
(\ref{cl2}).

The MDSA's discussion of both recoil effects 
and the Pauli principle is again highly misleading as it is based
exclusively on the Golden rule approximation and clearly
contradicts the energy-time uncertainty relation. Since MDSA use
the Pauli principle only as the energy constraint and argue that
electrons lose coherence ``by spontaneous emission'', their line
of reasoning is not at all specific to fermions and can equally be
applied, e.g., to a quantum particle interacting with
Caldeira-Leggett environment of harmonic oscillators.
Unfortunately MDSA avoid even mentioning about the exact solution
for the latter model which can be considered as a primer on zero
temperature quantum decoherence by interactions as well as a
demonstration of insufficiency of any Golden-rule-type
approximation for the problem in question.

As for Ref. \onlinecite{vD3} its main drawback lies in the
assumption of purely exponential decay of the Cooperon {\it at all
times} adopted by the authors. With this assumption any time
dependence slower than exponential would simply be excluded from
the analysis from the very beginning. The only contribution which
could be captured under this assumption is again the Golden rule
combination
$\coth\frac{\omega}{2T}+\tanh\frac{\epsilon-\omega}{2T}$. In
essence, all three approaches \cite{vD1,vD2,vD3} do not go beyond
exponentiating the first order Golden rule terms of the
perturbation theory. Therefore, it is not at all surprising that
all three approaches yield the same incorrect results for the
electron decoherence rate at $T=0$.

Despite all our critique the works \cite{vD1,vD2,vD3} have at
least one important merit. Namely, these authors have re-derived
our influence functional and acknowledged that our expressions (\ref{SR}), (\ref{SI})
are exact. This observation is crucial as it leaves practically no
room for further discussions. Moreover, it makes it easy also for
non-experts to judge which conclusion is correct without even
looking into complicated diagrams and path integrals. It is
actually sufficient to observe -- as JvD and MDSA do -- that both
$S_R$ and $S_I$ are {\it purely real}.

Consider two sums over $N$ realizations:
\begin{equation}
A=\frac{1}{N}\sum_{n=1}^{N}e^{-ia_n-b_n},
\;\;\;B=\frac{1}{N}\sum_{n=1}^{N}e^{i\alpha_n-ia_n-b_n},
\end{equation}
where $a_n$, $b_n$ and $\alpha_n$ are all real numbers. Obviously,
$B$ is transformed into $A$ by a trivial shift $a_n \to a_n +
\alpha_n$. Provided all $b_n$ are much larger than one, $b_n \gg
1$, both $A$ and $B$ are exponentially small,
$$A \sim B \sim \exp (-b), \;\;\; b={\rm min}_n b_n,$$
no matter what the values $a_n$ and $\alpha_n$ are. The same is,
of course, true for our path integral, one should only replace
realizations by trajectories and numbers by functionals $a_n \to
S_R$, $b_n \to S_I$.

Regrettably, these trivial algebraic considerations were not
respected by JvD. On p. 50 (very end of Sec. B5.8) he argues that
``this general argument would work if the measure used in the path
integral were real, however, it does not apply to the present case
''...``where the measure $e^{\pm iS_0}$ is complex''.  In other
words, JvD agrees that in the sum $A$ an imaginary part $ia_n$ has
no chance to cancel a real one $b_n$, but -- according to him --
this cancellation can happen in the sum $B$ where a complex
measure $e^{i\alpha_n}$ is added. JvD concludes ``Indeed, it is
shown in the main text''...``that contributions from $iS_R$ and
$S_I$ do partially cancel each other''.

Unfortunately it did not occur to JvD to make a shift $a_n \to a_n
+ \alpha_n$ in order to observe the full equivalence of $A$ and
$B$. Instead, he expands the exponent under the sum in $B$ to the
first order in $ia_n$ and $b_n$, observes that both contributions
$i\sum_ne^{i\alpha_n}a_n$ and $\sum_ne^{i\alpha_n}b_n$ may have
real parts which can (partially or exactly) cancel each other and
then re-exponentiates the result, i.e. writes
\begin{equation}
\tilde B \sim \exp
\left(-\frac{1}{N}\sum_{n=1}^{N}e^{i\alpha_n}(ia_n+b_n)\right).
\end{equation}
In this way JvD erroneously arrives at the ``cancellation'' in the
exponent. There is, of course, no need to analyze infinite series
of complicated diagrams in order to see that the quantity $\tilde
B$ has in general nothing to do with $B$, except in the
perturbative limit $a_n \ll 1$ and $b_n \ll 1$.

Finally, let us emphasize yet another point which we have already
discussed in Ref. \onlinecite{GZ4}. Very generally, evaluating a
non-Gaussian path integral around certain saddle point paths, at
$T \to 0$ one arrives at an effective action
\begin{equation}
S_{\rm eff}(t)=S^{\rm (cl)}(t) - \hbar \ln [A(\hbar ,t)],
\label{Scl}
\end{equation}
where $S^{\rm (cl)}(t)$ is the classical ($\hbar$-independent)
action on the relevant saddle point paths and $\hbar \ln A
(\hbar)$ represents the quantum correction ($A$ being the
pre-exponent). Obviously, this quantum correction can only be
important if $S^{\rm (cl)}(t) \lesssim \hbar$, i.e. in the
perturbative (short time) limit $t \ll \tau_\varphi$, in which
case partial cancellation of the term $S^{\rm (cl)}$ by the term
$\hbar \ln A (\hbar)$ is, of course, possible. However, for
$S^{\rm (cl)}\gg \hbar$ (i.e. for $t \gg \tau_\varphi$) there is
no way to cancel the classical action $S^{\rm (cl)}$ by the
quantum correction which formally tends to zero for $\hbar \to 0$.
Hence, in order to determine the scale $\tau_\varphi$ it is
absolutely sufficient to evaluate the action $-iS_R-S_I$ on pairs
of time-reversed saddle point paths, find the classical
($\hbar$-independent) action $S^{\rm (cl)}(t)$ and obtain the
decoherence time $\tau_\varphi$ from the condition $S^{\rm
(cl)}(\tau_\varphi )\sim \hbar$. This is exactly what was done in
our papers \cite{GZ2,GZ3,GZS}.

In conclusion, we have demonstrated that JvD's and MDSA's 
analysis \cite{vD1,vD2,vD3} fails
to correctly evaluate the low temperature decoherence rate for
electrons in disordered conductors.

\vspace{0.3cm}

 \centerline{\bf
Acknowledgment}

\vspace{0.3cm}

We are grateful to Jan von Delft for his persistent interest to
our work and for his openness with which he shared with us his doubts
on various points of our calculation.

\appendix

\section{}

For reference purposes we briefly recapitulate our path integral
representation for the conductance $\sigma$ of a disordered
conductor in the presence of electron-electron interactions. One
finds \cite{GZ2}
\begin{eqnarray}
\sigma =\frac{e^2}{3m} \int_{-\infty}^t dt' \int d{\bm r}_{Fi}d{\bm r}_{Bi}
\left(\nabla_{{\bm r}_{Ff}}-\nabla_{{\bm r}_{Bf}}\right)\big|_{{\bm r}_{Ff}={\bm r}_{Bf}}
\nonumber\\
 J(t,t';{\bm r}_{Ff},{\bm r}_{Bf};{\bm r}_{Fi},{\bm r}_{Bi})\big({\bm r}_{Fi}-{\bm r}_{Bi}\big)
\rho ({\bm r}_{Fi},{\bm r}_{Bi})
\label{sigma}
\end{eqnarray}
Here $\rho ({\bm r}_{Fi},{\bm r}_{Bi})$ is the equilibrium
single-electron density matrix, the function $J$ is given by the
path integral over the coordinates and momenta on the forward
(${\bm r}_F,{\bm p}_F$) and backward (${\bm r}_B,{\bm p}_B$)
branches of Keldysh contour,
\begin{eqnarray}
J&=&\int_{{\bm r}_F(t')={\bm r}_{Fi}}^{{\bm r}_F(t)={\bm
r}_{Ff}}{\cal D}{\bm r}_{F} \int_{{\bm r}_B(t')={\bm
r}_{Bi}}^{{\bm r}_B(t)={\bm r}_{Bf}}{\cal D}{\bm r}_{B} \int{\cal
D}{\bm p}_F\int{\cal D}{\bm p}_B \nonumber\\ &&\times\,
\exp\left\{iS_0[{\bm p}_F,{\bm r}_F]-iS_0[{\bm p}_B,{\bm
r}_B]\right\} \nonumber\\ &&\times\, \exp\left\{-iS_R[{\bm
p}_{F,B},{\bm r}_{F,B}]-S_I[{\bm r}_{F,B}]\right\}. \label{J}
\end{eqnarray}
The actions are defined as follows
\begin{eqnarray}
S_0[{\bm p},{\bm r}]&=&\int_{t'}^t dt'' \left({\bm p}\dot{\bm r}-\frac{{\bm p}^2}{2m}-U({\bm r})\right),
\label{S0}
\end{eqnarray}
\begin{eqnarray}
&& S_R=\frac{e^2}{2}\int_{t'}^t dt_1\int_{t'}^t dt_2
\label{SR}\\ &&\times\,
\bigg\{ R(t_1-t_2,{\bm r}_F(t_1)-{\bm r}_F(t_2))(1-2\rho({\bm p}_F,{\bm r}_F))
\nonumber\\ &&
-\,R(t_1-t_2,{\bm r}_B(t_1)-{\bm r}_B(t_2))(1-2\rho({\bm p}_B,{\bm r}_B))
\nonumber\\ &&
+\, R(t_1-t_2,{\bm r}_F(t_1)-{\bm r}_B(t_2))(1-2\rho({\bm p}_B,{\bm r}_B))
\nonumber\\ &&
-\, R(t_1-t_2,{\bm r}_B(t_1)-{\bm r}_F(t_2))(1-2\rho({\bm p}_F,{\bm r}_F))
\bigg\}, \nonumber
\end{eqnarray}
\begin{eqnarray}
 S_I&=&\frac{e^2}{2}\int_{t'}^t dt_1\int_{t'}^t dt_2
\big\{ I(t_1-t_2,{\bm r}_F(t_1)-{\bm r}_F(t_2))
\nonumber\\ &&
+\,I(t_1-t_2,{\bm r}_B(t_1)-{\bm r}_B(t_2))
\nonumber\\ &&
-\, I(t_1-t_2,{\bm r}_F(t_1)-{\bm r}_B(t_2))
\nonumber\\ &&
-\, I(t_1-t_2,{\bm r}_B(t_1)-{\bm r}_F(t_2))
\big\}. \label{SI}
\end{eqnarray}
Here $U({\bm r})$ is the impurity potential and $\rho({\bm p},{\bm r})$ is the electron density matrix. With sufficient
accuracy one can set $1-2\rho({\bm p},{\bm r})\simeq \tanh
[({\bm p}^2/2m +U({\bm r})-\mu)/2T]$.
The functions $R(t,{\bm r})$ and $I(t,{\bm r})$ read
\begin{eqnarray}
\label{RI}
R(t,{\bm r})&=&\int\frac{d\omega d^3q}{(2\pi)^4}\medskip
\frac{4\pi}{q^2\epsilon(\omega,q)}e^{-i\omega t+i{\bm qr}},
\\
I(t,{\bm r})&=&\int\frac{d\omega d^3q}{(2\pi)^4}\medskip
{\rm Im}\left(\frac{-4\pi}{q^2\epsilon(\omega,q)}\right)
\coth\bigg(\frac{\omega}{2T}\bigg)
e^{-i\omega t+i{\bm qr}},
\nonumber
\end{eqnarray}
where $\epsilon (\omega , k)$ is the dielectric function
\begin{equation}
\epsilon (\omega , q)=1+\frac{4\pi \sigma_D}{-i\omega +Dq^2} ,
\label{epsilon}
\end{equation}
$\sigma_D$ is the Drude conductivity and
$D$ is the diffusion coefficient.

For comparison, we also present the results obtained by JvD
\cite{vD1}, see Eqs. (1-4) of that paper. According to JvD the
conductivity should read
\begin{eqnarray}
\tilde \sigma&=&\frac{2e^2}{3m^2}\int dx_2 \left(\nabla_{{\bm
r}_1}-\nabla_{{\bm r}_{1'}}\right)\big|_{{\bm r}_1={\bm r}_{1'}}
\nonumber\\ &&\times\, \left(\nabla_{{\bm r}_2}-\nabla_{{\bm
r}_{2'}}\right)\big|_{{\bm r}_2={\bm r}_{2'}=x_2} \int
\frac{d\epsilon}{4T\cosh^2{\epsilon}/{2T}} \nonumber\\ &&\times\,
\int_0^\infty d\tau \tilde P^\epsilon(\tau; {\bm r}_1,{\bm
r}_{2'};{\bm r}_{1'},{\bm r}_{2} ), \label{sigmavD}
\end{eqnarray}
where
\begin{eqnarray}
&& P^\epsilon = \int_{{\bm r}_F(-\tau/2)={\bm r}_{2'}}^{{\bm r}_F(\tau/2)={\bm r}_{1}}{\cal D}{\bm r}_{F}
\int_{{\bm r}_B(-\tau/2)={\bm r}_{2}}^{{\bm r}_B(\tau/2)={\bm r}_{1'}}{\cal D}{\bm r}_{B}
\label{Pe}\\ &&\times\,
\exp\left\{i\tilde S_0[{\bm r}_F]-i\tilde S_0[{\bm r}_B]-i\tilde S_R[{\bm r}_{F,B}]-S_I[{\bm r}_{F,B}]\right\},
\nonumber
\end{eqnarray}
$\tilde S_0[{\bm r}]=\int_{-\tau/2}^{\tau/2} d t''\;( m\dot{\bm
r}^2/2-U({\bm r}))$ and
\begin{eqnarray}
\tilde S_R^{\epsilon}&=&\frac{e^2}{2}\int_{t'}^t dt_1\int_{t'}^t dt_2
\big\{ \tilde R_{\epsilon}(t_1-t_2,{\bm r}_F(t_1)-{\bm r}_F(t_2))
\nonumber\\ &&
+\,\tilde R_{-\epsilon}(t_1-t_2,{\bm r}_B(t_1)-{\bm r}_B(t_2))
\nonumber\\ &&
-\, \tilde R_{-\epsilon}(t_1-t_2,{\bm r}_F(t_1)-{\bm r}_B(t_2))
\nonumber\\ &&
-\, \tilde R_{\epsilon}(t_1-t_2,{\bm r}_B(t_1)-{\bm r}_F(t_2))
\big\}, \label{tildeSR}
\end{eqnarray}
where $\tilde R_{\epsilon}(t,{\bm r})=\tilde R_{\epsilon}^{Re}(t,{\bm r})+i\,\tilde R_{\epsilon}^{Im}(t,{\bm r}),$ 
\begin{eqnarray}
\tilde R_{\epsilon}^{Re}(t,{\bm r})=\frac{1}{2}\int\frac{d\omega d{\bm
q}}{(2\pi)^4}\,{\rm Re}\,\big[R(\omega,{\bm q})\big]\, e^{-i\omega t+i{\bm qr}} 
\label{reR}
\\  \times\,
\left[\tanh\frac{\epsilon-\omega}{2T} + \tanh\frac{\epsilon+\omega}{2T} \right],
\nonumber\\
\tilde R_{\epsilon}^{Im}(t,{\bm r})=\frac{1}{2}\int\frac{d\omega d{\bm
q}}{(2\pi)^4}\,{\rm Im}\,\big[R(\omega,{\bm q})\big]\, e^{-i\omega t+i{\bm qr}} 
\label{imR}
\\  \times\,
\left[\tanh\frac{\epsilon-\omega}{2T} - \tanh\frac{\epsilon+\omega}{2T} \right]
\nonumber
\end{eqnarray}
and $R(\omega,{\bm q})=\int dt d{\bm r}\, R(t,{\bm r})\,e^{i\omega
t-i{\bm qr}}$.


\begin{references}
\bibitem{FN1} Our derivation \cite{GZ2} is indeed a
formally exact procedure, as correctly stated by MDSA.  At the
same time the final expression for the influence functional
\cite{GZ2,GZ3,GZS} is tractable only within RPA.
\bibitem{FN3} This property is
general for the models in question and has nothing to do with the
presence or absence of $\rho$ in $S_R$.
\bibitem{FN4} Let us remind the reader that the importance of Hikami boxes
for the problem in question was repeatedly emphasized by other
authors \cite{AA} who made this point central for their critique
of our calculation. Now JvD, in contrast, argues that Hikami boxes
can be safely neglected in the first order. Here we do not need to
elaborate more on this controversy between JvD and the authors
\cite{AA} simply because the first order terms, even if fully and
correctly recovered, are still insufficient to draw any conclusion
about the electron decoherence time $\tau_\varphi$ which can only
be evaluated non-perturbatively, see, e.g. Refs.
\onlinecite{GZ3,GZS,GZ4} for further discussion.
\end{references}
\end{document}